\documentclass[11pt]{article}
\usepackage{graphicx}
\usepackage{amsmath}
\usepackage{amssymb}
\usepackage{amsxtra}

\usepackage{rotating} 
\usepackage{cite}

\title{DAMA/LIBRA results and perspectives}

\author{
R. Bernabei, P. Belli, S. d'Angelo, A. Di Marco, F. Montecchia\footnote{also Dip. di Ingegneria Civile e Ingegneria Informatica, Universit\`a di Roma ``Tor Vergata'', I-00133  Rome, Italy}\\
Dip. di Fisica, Univ. “Tor Vergata”\\
and INFN-Roma “Tor Vergata”, I-00133 Rome, Italy\\
F. Cappella, A. d'Angelo, A. Incicchitti\\
Dip. di Fisica, Univ. di Roma “La Sapienza”\\
and INFN-Roma, I-00185 Rome, Italy\\
V. Caracciolo\footnote{e-mail: vincenzo.caracciolo@lngs.infn.it}, R. Cerulli\\
Laboratori Nazionali del Gran Sasso, I.N.F.N., Assergi, Italy\\
C.J. Dai, H.L. He, X.H. Ma, X.D. Sheng, R.G. Wang, Z.P. Ye\footnote{also University of Jing Gangshan, Ji'an, Jiangxi, China}\\
Key Laboratory of Particle Astrophysics,\\ Institute of High Energy Physics,\\ Chinese Academy of Sciences, P.O. Box 918/3, Beijing 100049, China
}
\begin{document}
\maketitle

\begin{abstract}
The DAMA/LIBRA experiment ($\sim$ 250 kg of
highly radio-pure NaI(Tl)) is running deep underground at the
Gran Sasso National Laboratory (LNGS) of the I.N.F.N.
Here we briefly recall the results obtained in its first
phase of measurements
(DAMA/LIBRA--phase1; total exposure: 1.04 ton $\times$ yr).
DAMA/LIBRA--phase1 and the former DAMA/NaI
(cumulative exposure: $1.33$ ton $\times$ yr)
give evidence at 9.3 $\sigma$ C.L. for the presence of DM
particles in the galactic halo by exploiting the model-independent DM
annual modulation signature.
No systematic or side reaction able to mimic
the exploited DM signature has been found or suggested by anyone over more than a decade.
At present DAMA/LIBRA--phase2 is running with increased sensitivity.
\end{abstract}

\section{Introduction}

The DAMA project is based on the development and use of low
background scintillators. In particular, the second generation
DAMA/LIBRA apparatus \cite{perflibra,modlibra,modlibra2,modlibra3,bot11,pmts,mu,review,papep,cnc-l,IPP,dm14,results15,diu2014,norole,results,daq,cali,dm15,shadow,fase1-2},
as the former DAMA/NaI (see for example
Ref. \cite{review,RNC,ijmd} and references therein),
is further investigating the
presence of DM particles in the galactic halo
by exploiting the model independent DM annual modulation signature,
originally suggested in the mid 80's\cite{DrFr}.
At present DAMA/LIBRA is running in its phase2 with increased sensitivity.
The detailed description of the DAMA/LIBRA set-up during the phase1
has been discussed in details in Ref.~\cite{perflibra,modlibra,modlibra2,modlibra3,review,daq,cali,dm15,shadow,fase1-2}.

The signature exploited by DAMA/LIBRA (the model independent DM annual modulation)
is a consequence of the Earth's revolution around the Sun;
in fact, the Earth should be crossed
by a larger flux of DM particles around $\simeq$ 2 June
(when the projection of the Earth orbital velocity on the
Sun velocity with respect to the Galaxy is maximum)
and by a smaller one around $\simeq$ 2 December
(when the two velocities are opposite).
This DM annual modulation signature is very effective since the effect
induced by DM particles must simultaneously satisfy
many requirements: the rate must contain a component
modulated according to a cosine function (1) with
one year period (2) and a phase peaked roughly
$\simeq$ 2 June (3); this modulation must only be found in a
well-defined low energy range, where DM particle induced
events can be present (4); it must apply only to those events
in which just one detector of many actually ``fires'' ({\it single-hit}
 events), since the DM particle multi-interaction probability
is negligible (5); the modulation amplitude in the region
of maximal sensitivity must be $\simeq$  7\% for usually adopted
halo distributions (6), but it can be larger (even up to $\simeq$ 30\%)
in case of some
possible scenarios such as e.g. those in Ref.~\cite{Wei01,Fre04}.
Thus this signature is model independent, very discriminating and, in addition,
it allows the test of a large range of cross sections and of halo densities.
This DM signature might be mimicked only by systematic effects or side reactions
able to account for the whole observed modulation amplitude and
to simultaneously satisfy all the requirements given above.
No one is available \cite{perflibra,modlibra,modlibra2,modlibra3,mu,review,diu2014,Sist,RNC,ijmd,dm14,results15,norole,results,dm15,fase1-2}.

\section{The results of DAMA/LIBRA--phase1\\ and DAMA/NaI}

The total exposure of DAMA/LIBRA--phase1 is
1.04 ton $\times$ yr in seven annual cycles;
when including also that of the first generation DAMA/NaI experiment it is
$1.33$ ton $\times$ yr, corresponding to 14 annual cycles \cite{modlibra,modlibra2,modlibra3,review}.

\begin{figure*}[!ht]
\begin{center}
\includegraphics[width=0.95\textwidth] {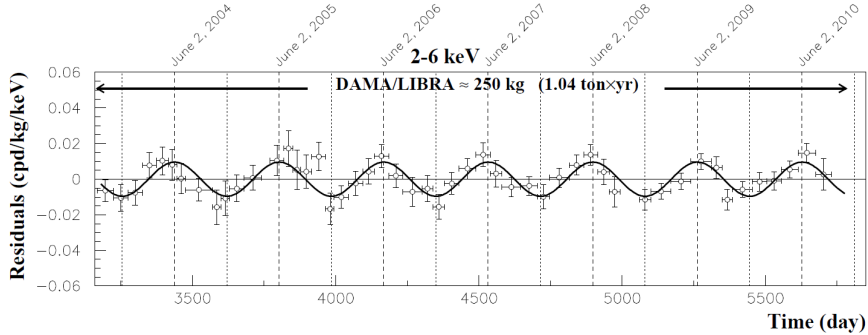}
\end{center}
\vspace{-.8cm}
\caption{Experimental residual rate of the {\it single-hit} scintillation events
measured by DAMA/LIBRA--phase1 in the (2--6) keV energy interval
as a function of the time.
The data points present the experimental errors as vertical bars and the associated
time bin width as horizontal bars.
The superimposed curves are the cosinusoidal functions behaviors $A \cos \omega(t-t_0)$
with a period $T = \frac{2\pi}{\omega} =  1$ yr, a phase $t_0 = 152.5$ day (June 2$^{nd}$) and
modulation amplitudes, $A$, equal to the central values obtained by best fit on these data points and those of DAMA/NaI.
The dashed vertical lines
correspond to the maximum expected for the DM signal (June 2$^{nd}$), while
the dotted vertical lines correspond to the minimum.}
\label{fg:res}
\end{figure*}

To point out the presence of the signal
the time behaviour of the experimental
residual rates of the {\it single-hit} scintillation
events for DAMA/NaI and DAMA/LIBRA--phase1 in the (2--6) keV energy interval
is plotted in  Fig.~\ref{fg:res}.
%
The $\chi^2$ test excludes the hypothesis of absence of modulation in the data:
$\chi^2$/d.o.f. = 83.1/50 for the (2--6) keV energy interval (P-value = 2.2 $\times$ 10$^{-3}$).
When fitting the {\it single-hit} residual rate of DAMA/LIBRA--phase1
together with the DAMA/NaI ones, with the function: $A \cos \omega(t-t_0)$, considering a
period $T = \frac{2\pi}{\omega} =  1$ yr and a phase $t_0 = 152.5$ day (June 2$^{nd}$) as
expected by the DM annual modulation signature,
the following modulation amplitude is obtained:
$A=(0.0110 \pm 0.0012)$ cpd/kg/keV corresponding to 9.2 $\sigma$ C.L..
When the period, and the phase are kept free in the fitting procedure,
the modulation amplitude is $(0.0112 \pm 0.0012)$ cpd/kg/keV (9.3 $\sigma$ C.L.),
the period $T= (0.998 \pm 0.002)$ year and the phase $t_0= (144 \pm 7)$ day,
values well in agreement with expectations for
a DM annual modulation signal. In particular, the phase is consistent
with about June $2^{nd}$ and is fully consistent with the value independently determined by Maximum Likelihood
analysis \cite{modlibra3}\footnote{
For completeness, we recall that a slight energy dependence of the phase
could be expected in case of possible contributions of non-thermalized DM components
to the galactic halo, such as e.g. the SagDEG stream \cite{epj06,Fre05,Gel01}
and the caustics \cite{caus}.}.
The run test and the $\chi^2$ test on the data
have shown that the modulation amplitudes singularly calculated for each
annual cycle of DAMA/NaI and DAMA/LIBRA--phase1
are normally fluctuating around their best fit values \cite{modlibra,modlibra2,modlibra3,review}.

We have also performed a power spectrum  analysis
of the {\it single-hit} residuals of DAMA/LIBRA--phase1 and DAMA/NaI
\cite{review}, obtaining a clear principal mode in the (2--6) keV energy interval at a frequency
of $2.737 \times 10^{-3}$ d$^{-1}$,
corresponding to a period of $\simeq$ 1 year, while only aliasing peaks are present just above.

Absence of any significant background modulation in the energy spectrum has been verified in
energy regions not of interest for DM \cite{modlibra3}; it is worth noting that the obtained results account of whatever
kind of background and, in addition, no background process able to mimic
the DM annual modulation signature (that is able to simultaneously satisfy
all the peculiarities of the signature and to account for the whole measured modulation amplitude)
is available (see also discussions e.g. in
Ref.~\cite{perflibra,modlibra,modlibra2,modlibra3,mu,review,diu2014,norole}).

A further relevant investigation in the DAMA/LIBRA--phase1 data has been performed by
applying the same hardware and software
procedures, used to acquire and to analyse the {\it single-hit} residual rate, to the
{\it multiple-hit} one. In fact, since the
probability that a DM particle interacts in more than one detector
is negligible, a DM signal can be present just in the {\it single-hit} residual rate.
Thus, the comparison of the results of the {\it single-hit} events with those of the  {\it
multiple-hit} ones corresponds practically to compare between them the cases of DM particles beam-on
and beam-off.
This procedure also allows an additional test of the background behaviour in the same energy interval
where the positive effect is observed.
In particular, the residual rates of the {\it single-hit} events measured in the (2--6) keV energy interval over
the DAMA/LIBRA--phase1 annual
cycles, as collected in a single cycle, are reported in Ref. \cite{modlibra3} together with the residual rates
of the {\it multiple-hit} events in the same energy interval.
A clear modulation satisfying all the peculiarities of the DM
annual modulation signature is present in the {\it single-hit} events, while
the fitted modulation amplitude for the {\it multiple-hit}
residual rate is well compatible with zero:
$-(0.0005\pm0.0004)$ cpd/kg/keV
in the same energy region (2--6) keV.
Thus, again evidence of annual modulation with the features required by the DM annual
modulation signature is present in the {\it single-hit} residuals (events class to which the
DM particle induced events belong), while it is absent in the {\it multiple-hit} residual
rate (event class to which only background events belong).
Similar results were also obtained for the last two annual cycles of the
DAMA/NaI experiment \cite{ijmd}.
Since the same identical hardware and the same identical software procedures have been used to
analyse the two classes of events, the obtained result offers an additional strong support for the
presence of a DM particle component in the galactic halo.

By performing a maximum-likelihood analysis of the {\it single-hit} scintillation events, it is possible to
extract from the data the modulation amplitude, $S_{m}$,
as a function of the energy considering $T=$1 yr and $t_0=$ 152.5 day.
Again the results have shown that positive signal is present in the (2--6) keV energy interval, while $S_{m}$
values compatible with zero are present just above; for details see Ref. \cite{modlibra3}.
Moreover, as described in Ref. \cite{modlibra,modlibra2,modlibra3,review},
the observed annual modulation effect is well distributed in all the 25 detectors,
the annual cycles and the energy bins at 95\% C.L.
Further analyses have been performed. All of them confirm the evidence for the presence
of an annual modulation in the data satisfying all the requirements for a DM signal.

Sometimes naive statements were put forwards as the fact that
in nature several phenomena may show some kind of periodicity.
The point is whether they might
mimic the annual modulation signature in DAMA/LIBRA (and former DAMA/NaI), i.e.~whether they
might be not only quantitatively able to account for the observed
modulation amplitude but also able to contemporaneously
satisfy all the requirements of the DM annual modulation signature. The same is also for side reactions.
This has already been deeply investigated in Ref.~\cite{perflibra,modlibra,modlibra2,modlibra3} and references
therein;
the arguments and the quantitative conclusions, presented there, also
apply to the entire DAMA/LIBRA--phase1 data. Additional arguments can be found
in Ref.~\cite{mu,review,diu2014,norole}.

No modulation has been found in any
possible source of systematics or side reactions; thus, cautious upper limits
on possible contributions to the DAMA/LIBRA measured modulation amplitude
are summarized in Ref. \cite{modlibra,modlibra2,modlibra3}.
It is worth noting that they do not quantitatively account for the
measured modulation amplitudes, and also are not able to simultaneously satisfy all the many requirements of the signature.
Similar analyses have also been done for the DAMA/NaI data \cite{RNC,ijmd}.
In particular, in Ref. \cite{norole} it is shone that, the muons and the
solar neutrinos cannot give any significant contribution to the DAMA annual modulation results.

In conclusion, DAMA give model-independent evidence (at 9.3$\sigma$ C.L. over 14 independent annual cycles)
for the presence of DM particles in the galactic halo.

As regards comparisons, we recall that no direct model independent comparison is possible in the field when different target materials
and/or approaches are used; the same is for the strongly model dependent indirect searches.
In particular, the DAMA model independent evidence
is compatible with a wide set of scenarios regarding the nature of the DM candidate
and related astrophysical, nuclear and particle Physics;
as examples
some given scenarios and parameters are discussed e.g. in
Ref.~\cite{RNC,modlibra,review} and references therein.
Further large literature is available on the topics.
In conclusion, both negative results and possible positive hints
are compatible with the DAMA model-independent DM annual
modulation results in various scenarios considering also the existing experimental and
theoretical uncertainties; the same holds for the strongly model dependent indirect approaches (see
e.g. arguments in Ref. \cite{review} and references therein).

The {\it single-hit}
low energy scintillation events
collected by \\DAMA/LIBRA--phase1  have also
been investigated in terms of possible diurnal effects\cite{diu2014}.
In particular, a diurnal effect with the sidereal time is expected for DM because of Earth rotation;
this DM second-order effect is model-independent and has several peculiar requirements as the DM annual modulation effect has.
At the present level of sensitivity the presence of any significant diurnal variation and of
diurnal time structures in the data can be excluded for both the cases of solar and sidereal time;
in particular, the DM diurnal modulation amplitude expected, because of the Earth diurnal motion,
on the basis of the DAMA DM annual modulation results is below the present sensitivity \cite{diu2014}.
It will be possible to investigate such a diurnal effect with adequate sensitivity
only when a much larger exposure will be available; moreover better sensitivities can also be achieved by lowering the software
energy threshold as in the presently running DAMA/LIBRA--phase2.

The data of DAMA/LIBRA--phase1 have also been used to investigate the so-called “Earth
Shadow Effect” which could be expected for DM candidate particles inducing nuclear recoils; this effect would be induced by the variation –
during the day – of the Earth thickness crossed by the DM particle in
order to reach the experimental set-up. It is worth noting that a similar
effect can be pointed out only for candidates with high cross-section with
ordinary matter, which implies low DM local density in order to fulfill
the DAMA/LIBRA DM annual modulation results. Such DM candidates
could get trapped in substantial quantities in the Earth’s core; in this
case they could annihilate and produce secondary particles (e.g. neutrinos) and/or they could carry thermal energy away from the core, giving
potentiality to further investigate them. The results, obtained by analysing in the framework of the Earth Shadow
Effect the DAMA/LIBRA--phase1 (total exposure $1.04$ ton$\times$yr) data are reported in Ref. \cite{shadow}.


For completeness we recall that other rare processes have also been searched for by
DAMA/LIBRA-phase1; see for details Refs. \cite{papep,cnc-l,IPP}.

\section{DAMA/LIBRA--phase2 and perspectives}\label{s:ph2}

An important upgrade has started at end of 2010 replacing all the PMTs with new ones having higher Quantum Efficiency;
details on the developments and on the reached performances in the operative conditions
are reported in Ref. \citen{pmts}. They have allowed us to lower the software energy threshold of the experiment
to 1 keV and to improve also other features as e.g. the energy resolution \cite{pmts}.

Since the fulfillment of this upgrade and after some optimization periods, DAMA/LIBRA--phase2
is continuously running in order e.g.:
\begin{enumerate}
 \item to increase the experimental sensitivity thanks to the lower software energy threshold;
 \item to improve the corollary investigation on the nature of the DM particle and related astrophysical, nuclear and particle physics arguments;
 \item to investigate other signal features and second order effects. This requires long and dedicated work for reliable collection and analysis of very large exposures.
\end{enumerate}

In the future DAMA/LIBRA will also continue its study on several other rare
processes as also the former DAMA/NaI apparatus did.

Moreover, the possibility of a pioneering experiment with anisotropic ZnWO$_4$
detectors to further investigate, with the directionality approach, those DM candidates that scatter off target nuclei is in progress \cite{adamo}.

Finally, future improvements of the DAMA/LIBRA set-up to further increase the sensitivity (possible DAMA/LIBRA-phase3) and the
developments towards the possible DAMA/1ton (1 ton full sensitive mass on the contrary of other kind of detectors), we proposed in 1996,
are considered at some extent. For the first case developments of new further radiopurer PMTs with high quantum efficiency are progressed,
while in the second case it would be necessary to overcome the present problems regarding: i) the supplying,
selection and purifications of a large number of high quality NaI and, mainly, TlI powders; ii) the
availability of equipments and competence for reliable measurements of small trace contaminants in ppt or lower region;
iii) the creation
of updated protocols for growing, handling
and maintaining the crystals; iv) the availability of large Kyropoulos equipments
with suitable platinum crucibles; v) etc.. At present, due to the change of rules for provisions of strategical materials,
the large costs and the lost of some equipments and competence also at industry level,
new developments of ultra-low-background NaI(Tl) detectors appear to be quite difficult.
On the other hand, generally larger masses do not imply a priori larger sensitivity;
in case the DM annual modulation signature is exploited, the improvement of other parameters
of the experimental set-up (as e.g. the energy threshold, the running time,...) plays an important role as well.

\end{document}